\documentclass[
               ]{jacow}

\usepackage[english]{babel}

\begin{document}

\title{\NoCaseChange{Hydrogen Electron Cyclotron Resonance Dissociator Demonstrator for the Project 8 Experiment}}

\author[1]{Manjinder Oueslati\thanks{mouesla@iu.edu}}
\author[1]{Walter Pettus}
\author[1]{Anna Reine}
\author[2]{Hamish Robertson}

\affil[1]{Center for Exploration of Energy and Matter, Indiana University, Bloomington IN, USA}
\affil[2]{Center for Experimental Nuclear Physics and Astrophysics, University of Washington, Seattle WA, USA}

\maketitle

\begin{abstract}
    Project 8, a neutrino mass experiment \cite{ref:P8Report}, employs the novel frequency technique of cyclotron radiation emission spectroscopy (CRES) to perform a model independent measurement of neutrino mass via tritium beta decay. Our pioneering experiment will consist of a beamline that dissociates molecular tritium, surface cools atoms through accommodation, evaporatively cools and slows atoms \cite{ref:MECB}, and traps them in a CRES resonant cavity to advance towards our sensitivity goal of \qty{40}{meV/c^2} for our final electron-weighted neutrino mass measurement.
    We are prototyping an electron cyclotron resonance (ECR) dissociator that uses a tunable \qty{1}{kW} \qty{2.45}{GHz} RF generator and  electromagnets to ignite a hydrogen plasma in an aluminum resonant cavity. Our ECR source is designed to produce a high atom flux ($\sim$\qty{e19}{atoms/second}) near room temperature. We present the engineering development, gas diagnostics, and spectroscopy data from our hydrogen plasma.
\end{abstract}
\section{Motivation}
\begin{figure}[!htb]
    \centering
    \includegraphics[width=1.0\linewidth]{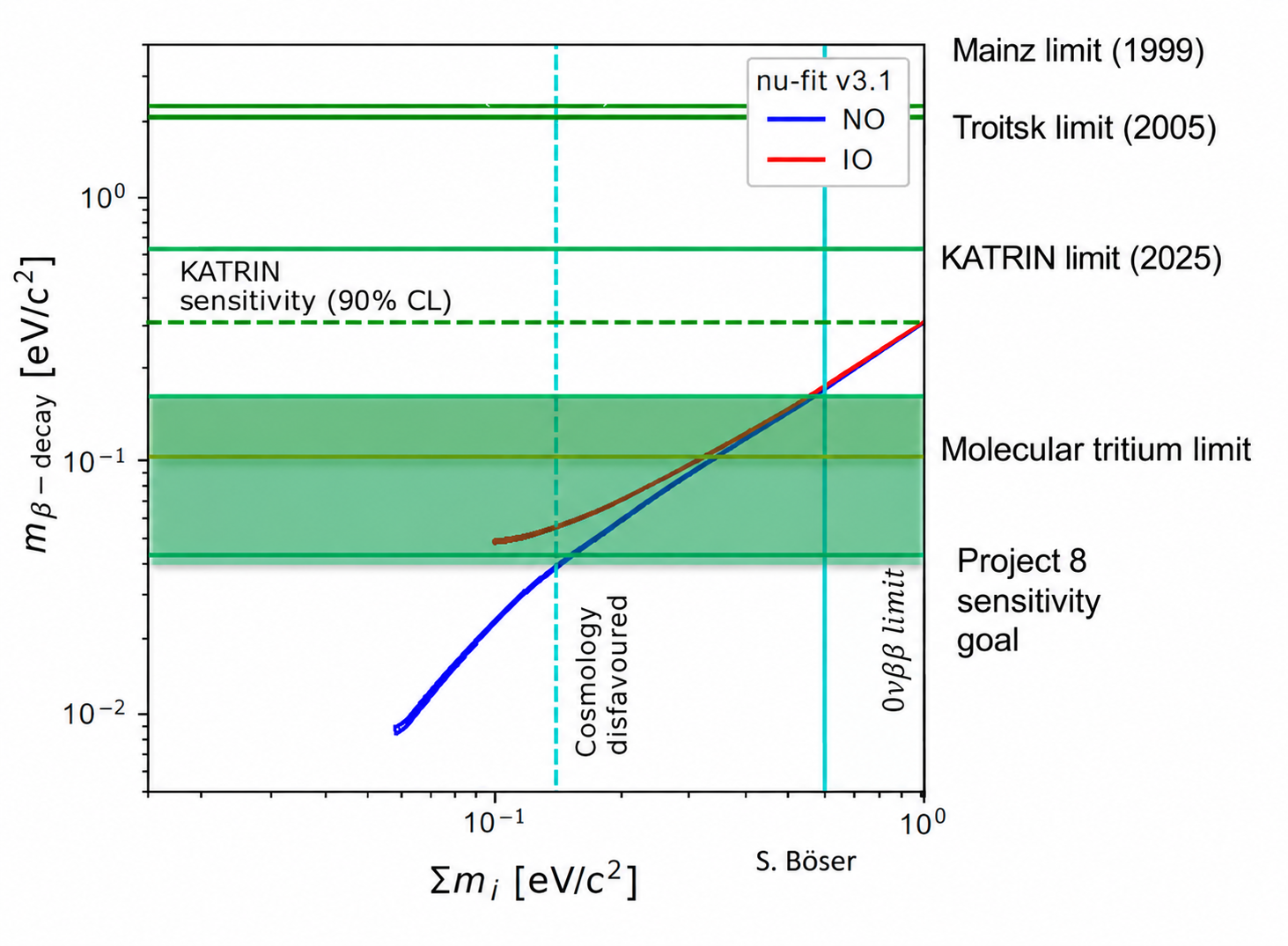}
    \caption{Electron-weighted neutrino mass limits from tritium beta decay with the inverted (normal) mass ordering shown in red (blue). Due to the statistical broadening on the molecular tritium endpoint, atomic tritium experiments are required to probe further into sensitive regimes. (Diagram is extrapolated from nu-fit parameters and annotated)}
    \label{fig:KATRIN}
\end{figure}

\begin{figure}[!htb]
    \centering
    \includegraphics[width=0.5\linewidth]{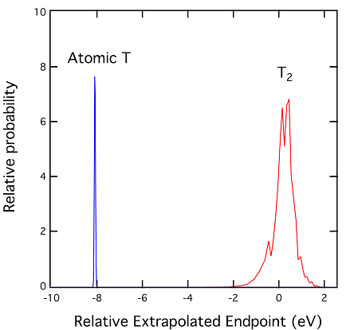}
    \caption{Difference in the extrapolated electron energy endpoint for atomic and molecular tritium with the broadening from the final states distribution of the tritium decay scheme \cite{ref:P8Report}.}
    \label{fig:Endpoint}
\end{figure}

    The Standard Model has seen unprecedented success in unraveling the fundamental nature of our universe. However, the first breach of its completeness arose when neutrinos were experimentally observed to oscillate between their flavor states, confirming a non-zero mass; unforeseen in the formation of the Standard Model \cite{ref:SNOLAB, ref:SuperK}. Since then neutrino oscillation experiments have sought to identify the mass ordering of the neutrino mass states, and physicists have pursued surveying the neutrino mass from particle flavor decay using model-independent kinematic experiments. At the forefront of mass searches, the KATRIN experiment \cite{ref:KATRIN2025} has made tremendous progress in measuring the elusive mass of the neutrino, setting the world-leading upper limit on the electron-weight neutrino mass from molecular tritium beta decay, \qty{450}{meV/c^2}, with a sensitivity of \qty{300}{meV/c^2} from remaining unreleased data. However, KATRIN is unable to go beyond their experimental sensitivity in Fig. \ref{fig:KATRIN} for two reasons: the statistical uncertainty associated with molecular tritium \cite{ref:Bodine2015} depicted in Fig. \ref{fig:Endpoint} and their integrated spectroscopic technique of magnetic adiabatic collimation with an electrostatic filter (MAC-E) to measure the kinetic energy of electrons with good precision.
    
\begin{figure}[!htb]
    \centering
    \includegraphics[width=1.0\linewidth]{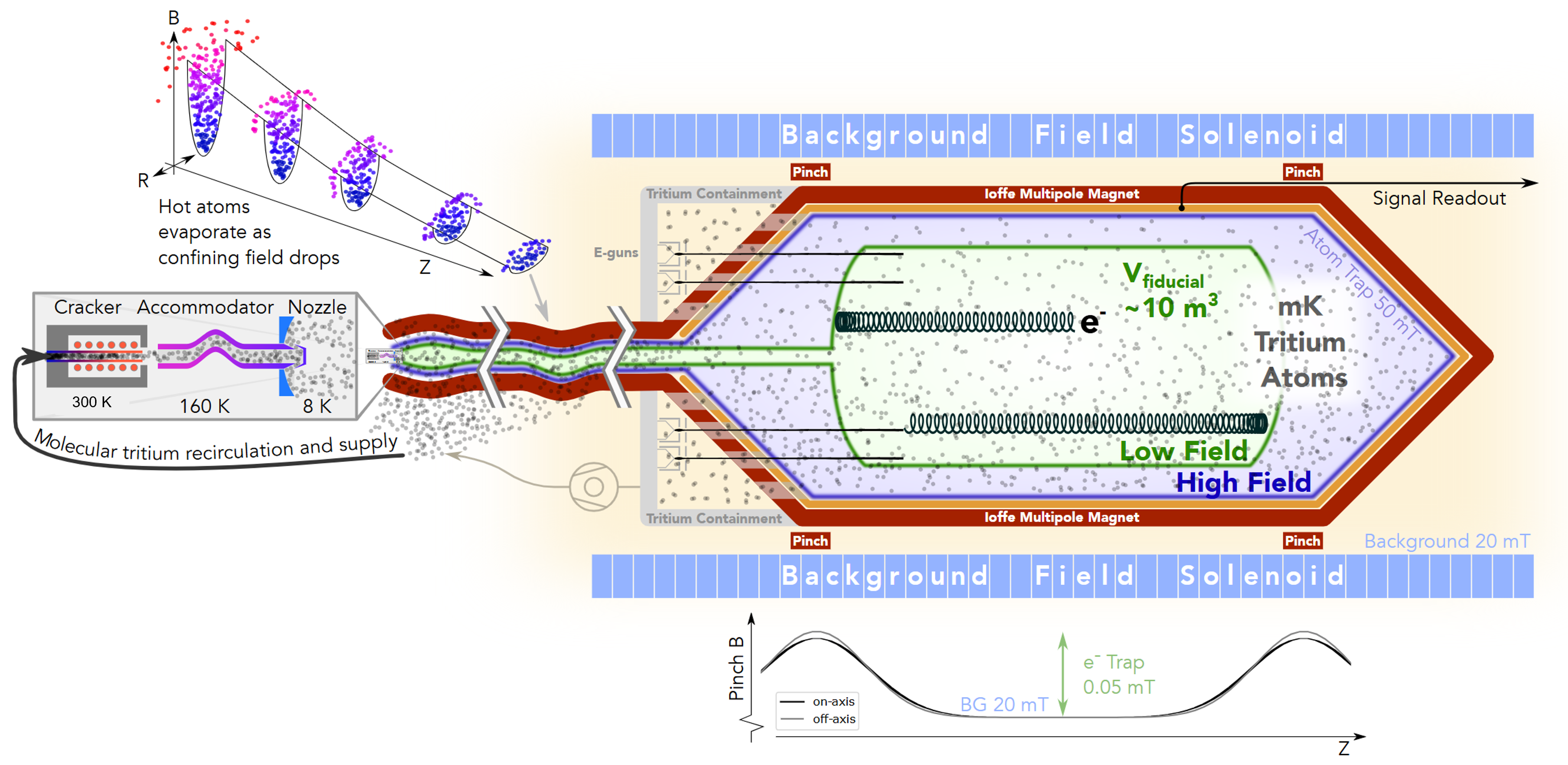}
    \caption{Project 8 is developing a high-flux atomic tritium source in which the atoms are magnetically cooled, guided, and state-selected to millikelvin temperatures before being confined within a large, high-order multipole magnetic trap. The trapped tritium atoms undergo beta decay, enabling measurement of the cyclotron frequency of the emitted electrons.}
    \label{fig:P8Concept}
\end{figure}

    Project 8 from its conception has been a frontier research R\&D atomic tritium experiment. To spearhead the next generation of neutrino mass experiments, Project 8 has developed a novel differential technique of cyclotron radiation emission spectroscopy (CRES) \cite{ref:CRES} that leverages directly measuring the natural cyclotron motion of an electron from tritium beta decay in a magnetic trap \cite{ref:P8PRL} with an experimental diagram in Fig. \ref{fig:P8Concept} to ultimately achieve a neutrino mass sensitivity of \qty{40}{meV/c^{2}}.

\section{Introduction}
Radio-frequency (RF) plasma sources were used in the original experiments that produced beams of atomic tritium to determine its hyperfine structure ~\cite{ref:Mathur1967}. 
All  experiments reported terminal inventory loss after only a few hours. This is partially understood to be the result of the attack of the plasma discharge from UV light, radicals, and fast electrons on the walls of the dissociator, which was made of glass or silica. We revisit the plasma source as a resonant cavity with a well-defined mode structure (TE111) that accelerates electrons and ions generally in the azimuthal direction, minimizing wall collisions. Ideally, the mode structure allows for a layer of unexcited gas separating the discharge region from the wall, by concentrating the fields and plasma on the radial center, with the expectation of protecting the chamber walls from sustained damage. The wall itself is not a dielectric, but aluminum. The good performance of aluminum is generally attributed to a native layer of oxide that forms on the surface, and the survival of this layer is important. It can be regenerated by exposure to oxygen, but that step would involve downtime and complexity and one goal of the R\&D is to measure and maximize uptime between regenerations. 

There is an optimal temperature range that minimizes atomic recombination at the chamber walls, which is crucial to maximize the outgoing atom flux and to maintain the purity of the source. At a temperature of about 160 K, tritium atoms can undergo many thousands of bounces before recombining. A temperature gradient can, moreover, be established from the discharge region to the exit to reduce the beam temperature, down to about 40 K or possibly below, with a thin-wall accommodator (exit tube) section.  

RF sources, and in particular electron cyclotron resonance (ECR) sources, display great success in producing relatively high beam, current outputs \cite{ref:Celona2014, ref:Saclay2000, ref:Slevin1981}. Moreover, some experiments involving new approaches to RF feeding have allowed an increase in the performance of conventional ECR ion sources with a minimum-B magnetic field structure by feeding with electromagnetic waves that have a large spectral content, or that are obtained by the superposition of a discrete set of microwaves at different frequencies. Although these experiments have provided interesting results, they have not provided us with an explanation or methodology to better understand the coupling mechanism between a feeding waveguide and a cavity filled with plasma, and the energy transfer between the electromagnetic field in the source plasma chamber and the plasma confined therein \cite{ref:Celona2014}. Studies of microwave coupling to the ECR plasma and therefore optimization of the ECR power transfer processes will allow us to design atom sources with higher performance.

Briefly, the degrees of freedom are: driven RF frequency (\qty{2.4}{GHz} -- \qty{2.5}{GHz}), RF generator power (\qty{20}{W} -- \qty{1000}{W}), cavity L/D (resonance mode), pressure, flow rate (\qty{0.1}{sccm} -- \qty{20}{sccm}), cavity material, cooling temperature of exit beam, (\qty{43}{K} -- \qty{300}{K}), accommodator L/D and magnet current (\qty{70}{A} -- \qty{110}{A}). We optimize these parameters to preserve the integrity of our apparatus with tritium contamination as well as to maximize our atom current output; to do that, we have developed a prototype ECR cavity.

\section{ECR experiment and design}
\begin{figure}[!htb]
    \centering
    \includegraphics[width=0.8\linewidth]{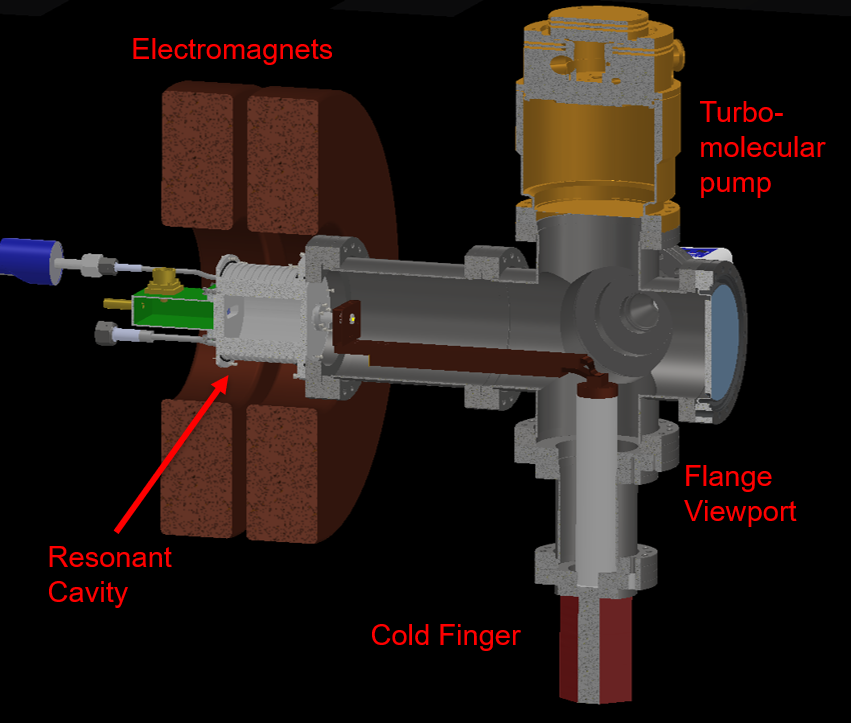}
    \includegraphics[width=0.9\linewidth]{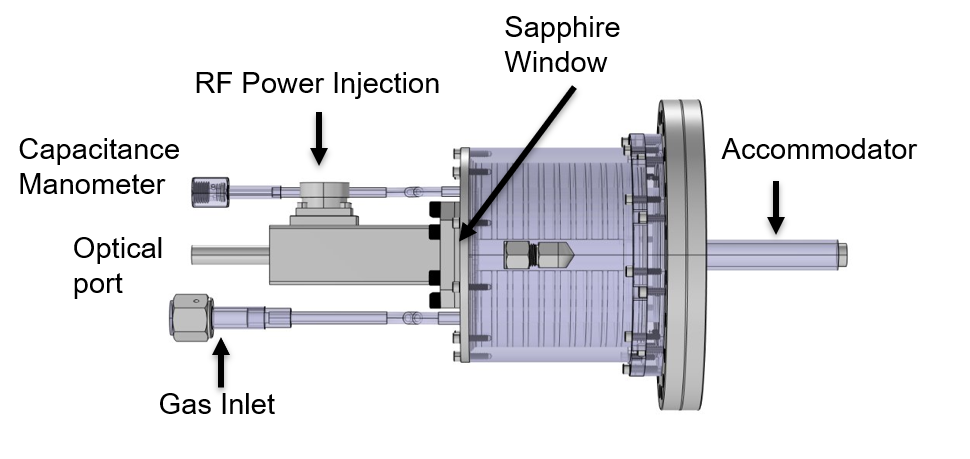}
    \caption{(Top) ECR hardware configuration. (Bottom) Detailed diagram of resonant cavity.}
    \label{fig:ECRSource}
\end{figure}
\begin{figure}[!htb]
    \centering
    \includegraphics[width=0.4\linewidth]{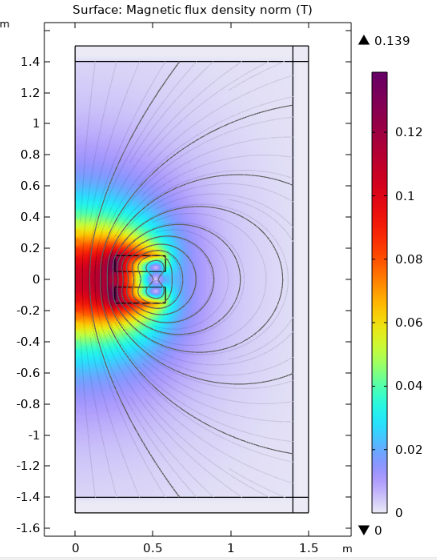}
    \includegraphics[width=0.55\linewidth]{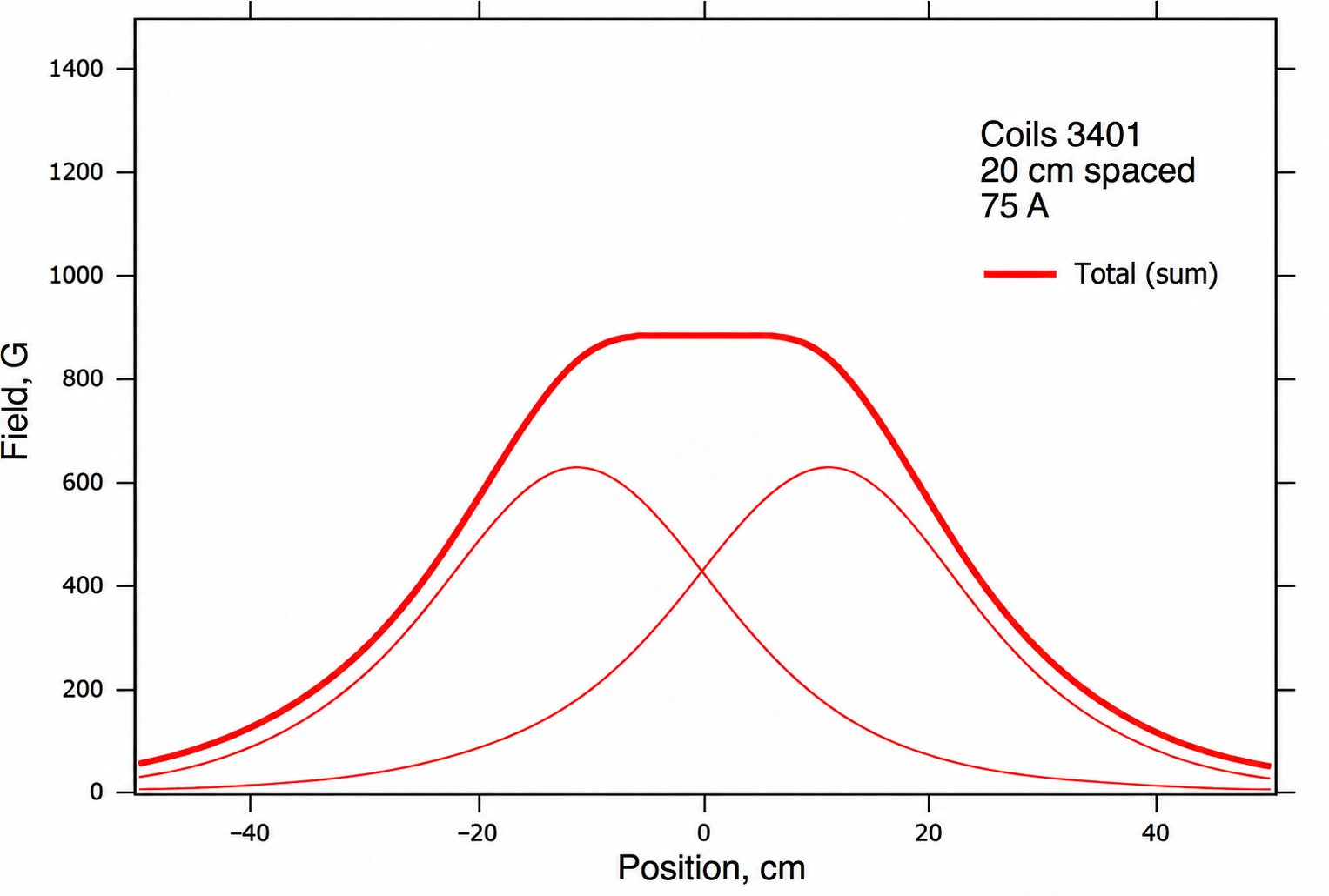}
    \caption{(Left) Magnetic fields COMSOL simulation of 2D magnetic flux profile for a homogeneous conducting coil geometry. (Right) Calculated axial magnetic field projection over the resonance region superimposed with data from measurements at a higher current of the magnet coils.}
    \label{fig:MagneticField}
\end{figure}
Neutral atoms can be produced in significant quantities within the plasma; however, the primary challenge is preserving the created atoms while maintaining an efficient extraction process from the source. This is where the accommodator plays a critical role. The residence time of the gas within the system is determined by the cavity pressure, the thermal velocity of atoms exiting the cavity, and the geometry of the accommodator. Together, these parameters set the relevant flow regime and therefore inform the modeling and optimization of atom transport and cooling through the accommodator. Operating at lower cavity pressures reduces the gas residence time and can limit opportunities for recombination, thereby improving the efficiency of the atom flux. To achieve the required low-pressure and low-temperature environment and efficiently remove the gas load, our system incorporates a molecular pumping system and cryogenic compressor. 

Our experimental apparatus in Fig. \ref{fig:ECRSource} consists of a turbo-molecular pump backed by a scroll pump with pressure gauges monitoring the exhaust, cross and cavity pressures. A cold finger to adjust the accommodator beam temperature profile. Electromagnets in an approximate Helmholtz configuration set the axial magnetic field in Fig. \ref{fig:MagneticField} for the resonance condition. RF power is injected into the cavity through a \qty{2}{m} RF cable, a WR284 transition stub, and a sapphire window ($\lambda_\textrm{RF}/4$ aperture depth to match the impedance of the cavity) to maximize RF transmission. A window to couple RF power is not compatible with a tritium system; however, it has demonstrated the capabilities and limitations of our hydrogen plasma source. Sapphire's properties consist of a low loss tangent making it an excellent dielectric material for high-frequency applications, high thermal conductivity for RF and plasma heating, and high melting point for extreme temperature and vacuum compatibility.
\begin{figure}[!htb]
    \centering
    \includegraphics[width=1.0\linewidth]{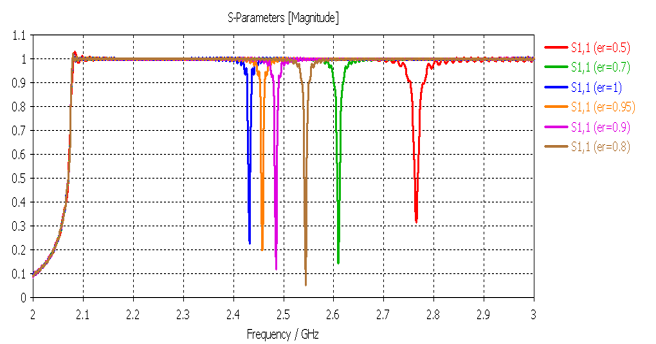}
    \caption{Simulation of plasma frequency shifts for varying dielectric plasma conditions. Small plasma shifts modeled as $\varepsilon_r < 1$, $\varepsilon_i \ll 1$.}
    \label{fig:FrequencyShift}
\end{figure}
The frequency shift of the resonant cavity upon plasma ignition is attributed to a change in the effective dielectric properties of the medium through which the incident RF wave propagates and validated by the Computer Simulation Technology Microwave and Electromagnetic (CST) solver in Fig. \ref{fig:FrequencyShift}. The resonant cavity was purposefully designed to be constructed at a natural frequency, \qty{2.43}{GHz}, and experimentally confirmed by a VNA adapter measured via oscilloscope, lower than the desired resonance frequency, \qty{2.45}{GHz}, for this reason. There is no simple solution for measuring the magnitude of the frequency shift; as the $S_{11}$ reflection coefficient could shift outside the tunable range of our RF generator.

\section{Diagnostics and Measurements}
\begin{figure}[!htb]
    \centering
    \includegraphics[width=1.0\linewidth]{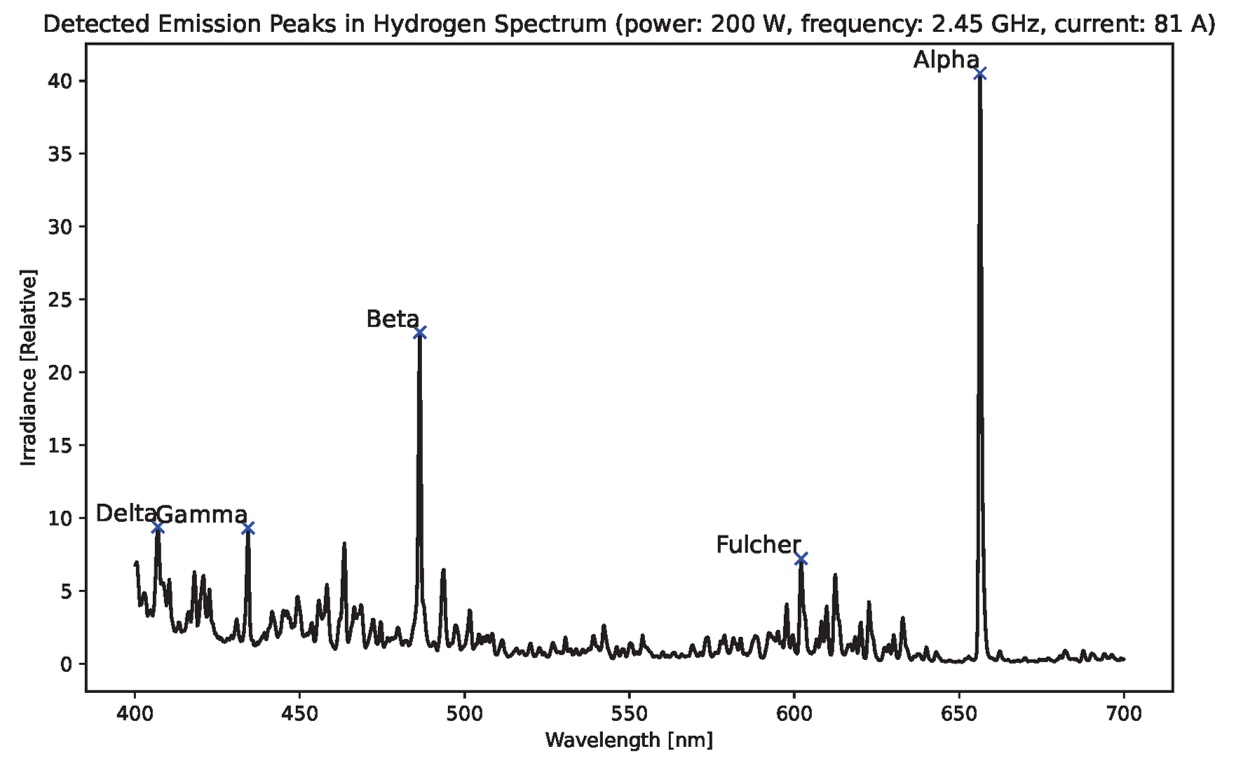}
    \includegraphics[width=0.4\linewidth]{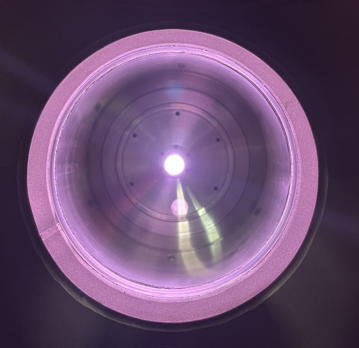}
    \caption{(Top) Blackbody calibrated hydrogen emission spectra scanned at \qty{10}{ms} of integration time averaged over 50 successive scans from OceanOptics SR4  spectrometer. (Bottom) Plasma visible to the eye through the flange view port and accommodator exit tube.}
    \label{fig:PlasmaSpectra}
\end{figure}
\begin{figure}[htb]
    \centering
    \includegraphics[width=0.5\linewidth]{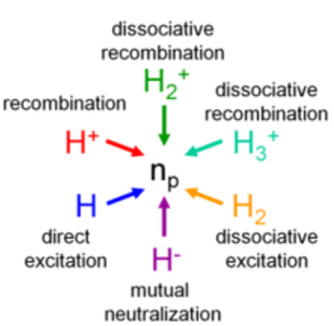}
    \caption{Reactive species channels that produce light emission of atomic and molecular hydrogen from \cite{ref:Fantz2016}.}
    \label{fig:ReactiveSpecies}
\end{figure}
Diagnostics performed both inside and outside the plasma provide key measurements for evaluating its performance as a dissociator. The sapphire window and an optical view port through the waveguide allowed us to perform Optical Emission Spectroscopy (OES), a key diagnostic in determining dissociation of molecular $H_2$ gas from excitation channels via collision radiative (CR) modeling. The objective is to determine the population densities of the excited states of specific atoms or molecules in the considered plasma, particularly the molecular Fulcher$-\alpha$ band and the Balmer atomic lines given in Fig. \ref{fig:PlasmaSpectra}. Our analysis relies on identifying the reactive species or excitation channels responsible for plasma excitations and the resulting spectral emissions, illustrated in Fig. \ref{fig:ReactiveSpecies}. To do this correctly, one must consider the intensity for a given emission from an excited state:
\begin{equation}
    I_{pq} = n_{0}n_{e}R_{0p}A_{pq}
\end{equation}
where $p$ is the excited principal quantum number ($\{{\alpha,\beta, \gamma, \delta}\} = \{{3,4,5,6}\}$) and $q$ is the de-excited quantum number (Balmer emission: $q=2$), $n_0$ represents the quasi-constant ground-state density of hydrogen, $n_e$ is the plasma electron density, $A_{pq}$ is the Einstein coefficient that describes the probabilities of radiative transitions between two energy levels for photon emission and absorption, and $R_{0p}$ is the population coefficient that describes how a particular excited state depends on the plasma parameters and is calculated from a CR model (Yacora on the Web) considering the differential processes that populate or depopulate the excited level \cite{ref:Yacora}. Prior to determining a particle fraction, we must ascertain the plasma parameters within our experiment. We do so indirectly: the electron density is computed by an analytic calculation from our driven RF power $n_e \sim$ \qty{e17}{m^{-3}} and the electron temperature $T_e$ is obtained from the relative intensity ratio of 2 atomic lines (assuming a Maxwellian electron energy distribution function (EEDF) electron energies in the CR model):
\begin{equation}
    \frac{I_\beta}{I_\gamma} = \frac{n_e n^H_0\textcolor{red}{{R_{0\rightarrow4}}}A_{4\rightarrow2}+...}{n_e n^H_0\textcolor{red}{{R_{0\rightarrow5}}}A_{5\rightarrow2}+...}
\end{equation}
\begin{figure}[htb]
    \centering
    \includegraphics[width=0.75\linewidth]{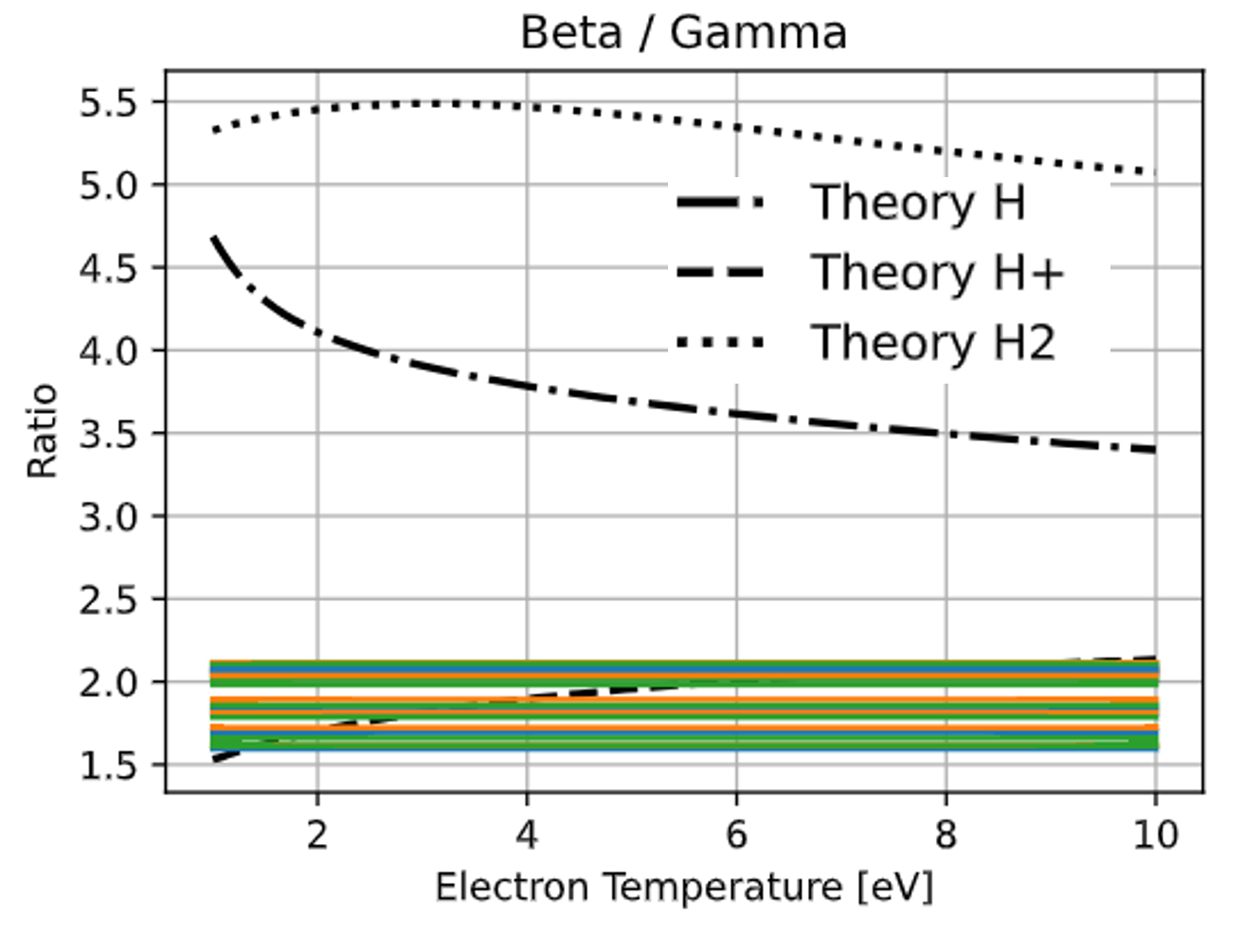}
    \caption{Theoretical line ratio and data ratio for the $H_\beta$ and $H_\gamma$ Balmer emission lines. These lines are generally considered to significantly contribute to the atom species \cite{ref:Verhaegh,ref:Mazzaglia,ref:Deng}.}
    \label{fig:ElectronTemperature}
\end{figure}
The same process can be performed to evaluate the particle fraction using a ratio of the atomic emission lines to the molecular Fulcher$-\alpha$ band:
\begin{equation}
    \frac{I_\beta, I_\gamma}{I_{\textrm{Ful}-\alpha}} = \frac{n_e \textcolor{red}{n^H_0}{R_{0\rightarrow4,5}}A_{4,5\rightarrow2}+...}{n_e \textcolor{red}{n^{H_2}_0}{R_{0\rightarrow d^3}}A_{d^3\rightarrow a^3}}
\end{equation}
We were unable to calculate a particle fraction from our data. In calibrating our spectrometer for plasma intensity, we had applied a relative or blackbody calibration that corrects our raw spectrum acquisition into radiation intensity by accounting for the instrument's spectral response, skewed measurements by grating efficiency, fiber-optic cable losses, and detector sensitivity variations, which deviate significantly from the ideal blackbody behavior of Planck's Law. When we computed the ratios, we assumed that a single excitation channel dictates the interactions and mechanisms within the plasma at any given moment. However, that is an oversimplification of plasma dynamics and thus insufficient calibration for our spectrometer.

Another diagnostic to classify our atom current is a rise in the pressure inside the cavity from molecular dissociation. The accommodator provides a means to measure the dissociation because the pumping speed through the tube is proportional to $1/\sqrt{M}$, with $M$ being the input of a constant molecular mass flow regulated by a mass flow controller. If dissociation does not occur, the density in the cavity is determined by the inflow rate and the flow impedance of the accommodator. If dissociation is complete, then the thermal velocity is greater by $\sqrt{2}$ and the number of atoms escaping is doubled, but the mass flow is unchanged because it is set by the molecular inflow. Since the escape speed is increased, the density must be smaller to keep the mass flow the same. During monitoring the plasma chamber pressure during plasma ignition, we observed a gradual pressure drift over time. After the plasma was extinguished, the chamber pressure did not return to its initial base pressure. This behavior suggests that plasma operation induces a hysteresis effect in the chamber pressure gauge.

\section{Experimental Outlook}
ECR plasmas typically have relaxed conditions in generating the plasma, the difficult portion comes when we attempt to study and understand the dynamics and interactions that govern the plasma. As such upgrading our diagnostic tools is at the forefront of our future. Improving our OES diagnostic with an absolute calibration source (Ulbricht sphere or tungsten lamp) to adequately determine the radiant flux of our plasma is paramount to consider the geometric efficiencies of our light collection method as well as perform a rigorous treatment of the reactive species in the plasma. Another spectroscopic study is to assess source contamination via a helium plasma. Helium is an inert gas and lacks a molecular counterpart, thus making it an excellent candidate to calibrate our pressure drift measurements.

The method of pressure dissociation depends on the atoms flowing out of the accommodator, remaining atoms rather than recombination prevailing. At room temperature, some recombination will occur in the accommodator, but lowering the temperature of the accommodator tube to approximately 160 K the recombination rate should be minimal and the pressure diagnostic will improve. However, when the accommodator is cooled, the flow impedance for both atoms and molecules increases, so we have to measure the cavity pressure in the absence of a discharge and with the discharge igniting. We will be able to tell if thermal accommodation is working because we should see the cavity pressure increase as $\sqrt{T}$ for the temperature of the exit tube. The temperature of the gas flowing through the accommodator is
expected to be close to that of the tube itself, because thermal accommodation occurs quickly. We have installed a cold finger to complete this upgrade.

\section{Conclusion}
We have developed a prototype to test our assumptions regarding plasma ignition and sustainability to prepare for radioactive tritium. ECR plasmas have demonstrated success in producing high-intensity ion currents, making them a promising candidate as an atomic source for our experiment. One of the primary challenges surrounding their implementation, however, is the longevity of plasma-facing surfaces. Oxide-coated aluminum walls have demonstrated robustness in comparison with previous chamber designs, which have shown limited success in sustaining a stable plasma over extended periods. Ideally, the cavity mode structure should help minimize plasma-surface interactions by concentrating the electromagnetic field away from the chamber walls. Although the electron mean free path is relatively short, this does not preclude energy deposition at the walls. Continued energy deposition can contribute to surface degradation and material damage over extended periods of plasma operation. Therefore, material selection plays an important role, as the thermal conductivity of the plasma surface governs heat transfer during wall collisions and influences the thermal accommodation of the atoms. Additionally, our external cooling of the accommodator to further reduce the temperature promotes stabilization and minimizes the recombination of the outgoing atom population. Variations in accommodator design will require retuning ECR parameters if plasma stability continues to be a problem.

Accurately measuring and characterizing the plasma currents remains the largest experimental hurdle. However, there is a clear path forward through the continued development and improvement of our diagnostics, with the ultimate goal of reliably measuring the extracted atom beam flux.

\section{Acknowledgments}
We thank Tianhuan Luo of Lawrence Berkeley Lab for the cavity calculations. This material is based upon work supported by the following sources: the U.S. Department of Energy Office of Science, Office of Nuclear Physics, under Award No.~DE-FG02-97ER41020 to the University of Washington. The National Science Foundation under Grant No.~PHY-2209530 and PHY-2609888 to Indiana University. This work is made possible through the efforts of the Project 8 collaboration.


\end{document}